\documentclass[aps,prl,preprint,showpacs,showkeys,letterpaper,floatfix]{revtex4}

\usepackage[dvips]{graphicx}

\begin{document}

\title{Decaying Higgs Fields and Cosmological Dark Energy}

\author{Robert J. Nemiroff and Bijunath Patla}

\affiliation{Michigan Technological University, Department of Physics, \\
1400 Townsend Drive, Houghton, MI  49931}

\date{\today}

\begin{abstract} 
The observed dark energy in the universe might give particles inertial mass.  We investigate one realization of this idea, that the dark energy field might be a decayed scalar component of a supermultiplet field in the early universe that creates inertial mass through spontaneous symmetry breaking, e.g. a Higgs field.  To investigate this possibility, the cosmological Friedmann equation of energy balance is augmented in a standard way to incorporate a minimally coupled cosmological Higgs. For epochs where the expansion of the universe is driven by matter and radiation and not the scalar field, the observed hidden nature of the Higgs field can be codified into a single differential equation that we call the ``hidden higgs" condition.  The resulting differential equation is solved for the time dependant scalar field and a simple and interesting solution is found analytically.  Such a Higgs field decays from Planck scale energies rapidly and approximately exponentially from onset, leaving only the initially negligible constant term of the potential as a final cosmological constant.  Such evolution replaces the hierarchy problem with the problem of explaining why such evolution is physically justified. 
\end{abstract}
\maketitle

\section*{Introduction}
Scalar Higgs fields are expected to exist in our universe in order to create inertial mass \cite{Englert}.  Yet, these fields have never been uniquely detected.  Conversely, dark energy was not expected to exist in our present universe, yet has been uniquely detected.  This paper is an attempt to examine the possibility that the unexpected dark energy might actually be related to the expected Higgs fields, possibly even the electroweak Higgs field.  Attributing the dark energy to a cosmological Higgs field is satisfying in at least one other philosophical context: the dark energy in the universe might then be {\it needed} to give the universe inertial mass.

The well known ``hierarchy problem" might appear to make known scalar fields such as the Higgs field inappropriate for a cosmological setting \cite{Veltman74}.  Briefly, the hierarchy problem states that the canonical energy density on particle physics scales are many orders of magnitudes higher than those detected or limited in a cosmological setting. The hierarchy problem is particularly evident when comparing the value of observed cosmological dark energy to the electroweak energy scales.  We suggest here a dynamical solution of a decaying scalar field, allowing a large scalar field in the early universe to evolve into the small scalar field seen today.  Although this solution invokes a force to move the scalar field, we feel that postulating such a force is still preferable to the alternative of just assuming that Higgs fields abruptly no longer exist in a cosmological setting.

Using Higgs fields in a cosmological setting was first tried in the context of non-adiabatic expansion in the early universe \cite{Kolb80} and early models of inflation \cite{guth}.  A recent attempt \cite{vanHolten02} to find a solution for a universe dominated by a Higgs field suggested that a cosmological Higgs field might bridge the gap between the Planck and GUT scales. Inspired by the possibility that the dark energy is the only presently necessary scalar field, we decided to extend van Holten's work to the present epoch.

The cosmological Higgs field discussed here might be a component of a supermultiplet scalar field.  If so, an isodoublet responsible for the electroweak sector which has a scalar component left unabsorbed by the guage bosons and fermions should also have been a part of the supermultiplet in the early universe \cite{spira}.  This field is likely different than the inflaton field which might have played a role in trapping the electroweak Higgs field during reheating \cite{Lyth}. So we have every reason to believe that the {\it cosmological} Higgs field is not the {\it inflaton} field which is responsible for inflation and only related to particle {\it electroweak} Higgs field that gives the Higgs particle its mass.  We are curious to know if a time-dependant cosmological Higgs field could be consistent, even if only formally, with recent measurements detecting and limiting the dark energy.  It may or may not be the field that gives {\it dark} matter its mass.

\section*{Cosmological Hidden Higgs Scenario}

Starting with a flat Friedmann-Robertson-Walker-type universe, we add the dynamic energy density of the component of the scalar field which also obeys the Klein-Gordon equation of motion,  ${\ddot \phi} + 3 H {\dot \phi} + dV/d{\phi} = 0$ to the Friedmann's equation:
\begin{equation}
 {1 \over 2} {\dot \phi}^2 + V(\phi) +  \rho_m + \rho_r = \rho_{crit} ,
\label{FRW1}
\end{equation}
where $\phi$ is a minimally coupled Cosmological Higgs field, $V(\phi)$ is the potential of the field,  $\rho_m$ and $\rho_r$ are the density in matter and radiation, and $\rho_{crit}$ is the critical mass-energy density that makes the universe spatially flat.  Now $\rho_{crit} = 3 H^2/(8 \pi G)$ where $H$ is the time dependant Hubble parameter and $G$ is the non-time-dependant gravitational constant.
\begin{equation}
 V(\phi) = {\lambda \over 4} \phi^4  - {\mu^2 \over 2} \phi^2 + \epsilon ,
\label{HiggsV}
\end{equation}
where $\lambda < 1$, is a dimensionless coupling constant, $\mu^2$ and $\epsilon$ are positive constants with the units of $GeV^{2}$ and $GeV^{4}$ respectively. $\lambda$ and $\mu$ are commonly assumed to be constant throughout time \cite{Linde90}.  $\epsilon$, discussed in more detail later, is the vacuum expectation value of the cosmological Higgs field, which will become today's dark energy. We further note that $\rho_m \sim \rho_{mo} / a^3$ and $\rho_r \sim \rho_{ro}/ a^4$ where $\rho_{mo}$ and $\rho_{ro}$ are the values of the matter and radiaton density at some time $t_o$ and $a = a(t)$ is the time dependant scale factor of the universe relative to the present.  Note that $1/a(t) = (1+z)$ and that $H = {\dot a}/a$.  Equation (\ref{FRW1}) then becomes
\begin{equation}
 {1 \over 2} {\dot \phi}^2 + 
 {\lambda \over 4} \phi^4  - {\mu^2 \over 2} \phi^2 + \epsilon
 + {\rho_{mo} \over a^3} + {\rho_{ro} \over a^4} 
 = {3 {\dot a}^2 \over 8 \pi G a^2} .
\label{FRW2}
\end{equation}
Now this differential equation convolves two time-dependant functions: $\phi$ and $a$ (and their time derivatives).  In general, it is impossible to solve for them both.  We can assume, however, that $a$ and hence ${\dot a}$ is measured or at least constrained in the recent universe by observations.  Inputting $a$, it then becomes possible to solve for the time behavior of $\phi$.

There exists a simple solution for Eq. (\ref{FRW2}), when matter and radiation dominate the universe. The terms with explicit $\phi$ and ${\dot \phi}$ dependence might not be needed to determine $a$. Mathematically speaking:
\begin{equation}
 {1 \over 2} {\dot \phi}^2 +  {\lambda \over 4} \phi^4  - 
 {\mu^2 \over 2} \phi^2 = 0 .
\label{HiddenHiggs}
\end{equation}
which, we shall call ``The Hidden Higgs" condition. Note that this equation is only an approximation. In general we can only say that the terms on the left hand side collectively are very small compared to the remaining terms on the left hand side of Eq. (\ref{FRW2}). This leads to two cases: either $\phi$ and ${\dot \phi}$ are always small individually or there is some cancellation involved and so they are small collectively.  If they are each individually small, then their solution is likely lost to the inaccuracy of our knowledge of the time dependence of $a$, particularly at large redshifts.  This also leaves the question of how they got so small unanswered.

If the terms are collectively small, however, then there are again two cases.  In the first case ${\dot \phi}$ is individually small and the two $\phi$ terms are collectively small.  The only way this can happen is if ${\lambda \over 4} \phi^4 \sim {\mu^2 \over 2} \phi^2$.  In this case the simple solution is $\phi^2 \sim 2 \mu^2 / \lambda$. 

In the second case, terms only cancel collectively through the general solution of the Eq. (\ref{HiddenHiggs}).  When the approximate equality is exact, this differential equation can be written as the integral 
\begin{equation}
 \int_{t_{initial}}^{t_{final}} \ dt = \pm
 \int_{\phi_{initial}}^{\phi_{final}} 
 { 2 \ d\phi \over \sqrt{ 2 \mu^2 \phi^2 - \lambda \phi^4 } } .
\end{equation}
This integral, and hence differential Eq. (\ref{HiddenHiggs}), has analytic solutions.  We choose the solution with $\phi$ is always positive and decreasing toward zero at infinite time.  We take the initial time as $t = 0$ where $\phi^2 = 2 \mu^2 / \lambda$. Then
\begin{equation}
 t = { 1 \over \mu }
 {\rm ln} \left( 
 { \sqrt{ 2 \mu^2 / \lambda} + \sqrt{ 2 \mu^2 / \lambda - \phi^2 }
 \over \phi } \right) .
\label{tanalytic}
\end{equation}
The term $\epsilon$ in Eq. (\ref{HiggsV}) is not part of the mathematical Hidden Higgs condition and so survives.  Indeed this is by design so that it can become the asymptotic value of the dark energy. The time dependant nature of the Higgs field is only expected to evolve, at the earliest, from the start time corresponding to the initial value of $\phi _{in} = {\sqrt{2}{\mu }}/\sqrt{\lambda }$. Actually, the time evolution might start later when Eq. (\ref{HiddenHiggs}) becomes valid, which might correspond to a time when the Higgs field no longer dominated the energy density of the universe.  Therefore, this solution can only be valid in the range $0 < \phi < \sqrt{2}\mu / \sqrt{\lambda}$ and that $\phi$ is near zero then $t$ is near infinite.
Eq. (\ref{tanalytic}) can be inverted and rewritten as
\begin{equation}
 \phi = { \sqrt{2 \mu^2 / \lambda} e^{-\mu t} \over
          1 + e^{-2 \mu t} } .
\label{phianalytic}
\end{equation}

\begin{figure}
\includegraphics[width=0.87\linewidth, height=0.55\linewidth]{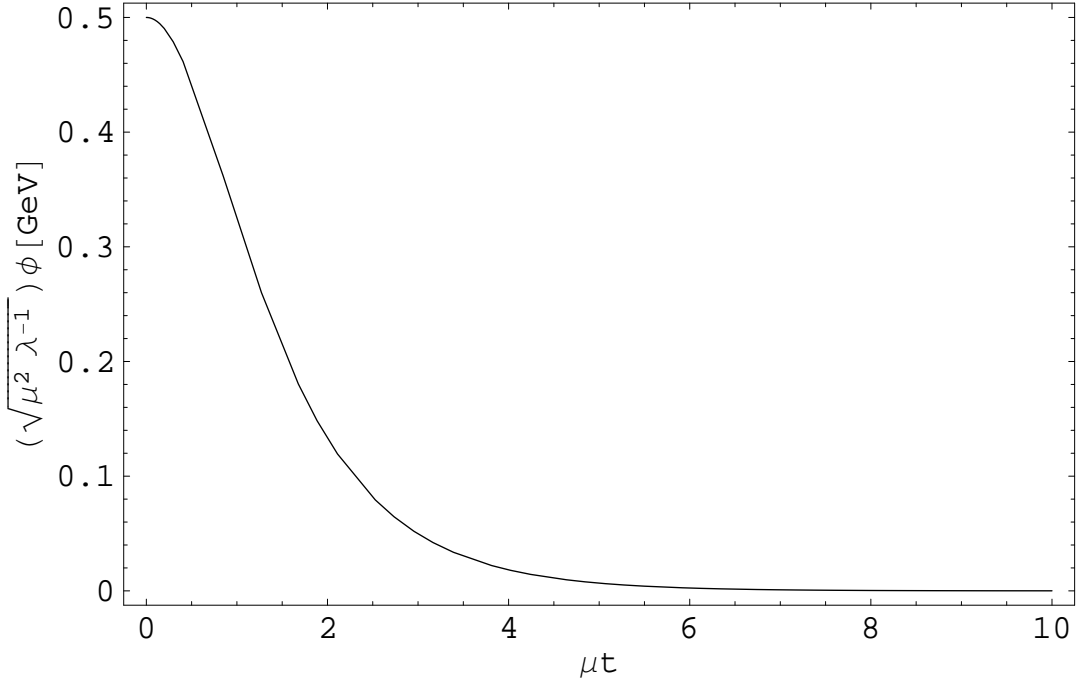}
\caption{The decaying Higgs field }
\label{fig1}
\end{figure}

A plot of $\phi$, in units of $\sqrt{ \mu^2 / \lambda}$, versus time is given in Fig.~\ref{fig1}.  Note that the choice of $\mu$ uniquely scales the time coordinate.  At long times, considered to be when $t >> 1 / \mu$ and $\phi << \sqrt{2 \mu^2 / \lambda}$, Eq. (\ref{phianalytic}) becomes a simple exponential and can be written
\begin{equation}
 \phi \sim \sqrt{2 \mu^2 \over \lambda} e^{-\mu t} .
\end{equation}
This indicates that at late times the non-constant parts of the scalar field decay exponentially, eventually leaving only the constant $\epsilon$ part.  Such behavior is not uncommon for certain types of scalar fields, including a Cosmological Higgs field when it dominates the universe energy density \cite{vanHolten02}.

\section*{Results and Discussion}

\begin{figure}[htb]
\includegraphics[width=0.87\linewidth, height=0.55\linewidth]{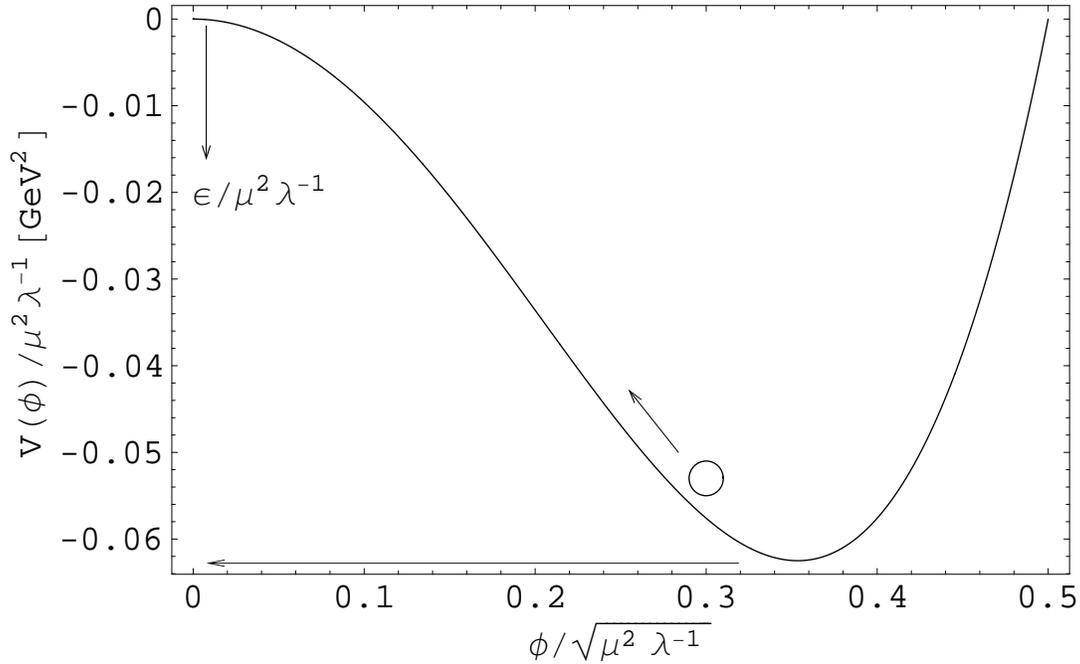}
\caption{The Cosmological Higgs Potential for the field climbing uphill after rolling down.}
\label{fig2}
\end{figure}

The initial value of $\phi$ depends on $\mu $ and $\lambda$. We'll assume ${\mu}^2 = M^4/{M_{P}^{2}}$, where M is some intermediate scale, possibly of the order SUSY breaking scale, where ${M_{P}} \approx 10^{19}$ GeV, the Planck's energy. The minimum of the potential $V(\phi)$ in Fig~\ref{fig2} correspond to $\mu t \approx 1$, which happens to be the Planckian regime and the Cosmological Higgs field $\phi$ attains the value  $\sqrt { \mu^2 / 4 \lambda}$. We presume that SUSY is broken before $\phi$ decays.

It is now possible for us to formulate the exact expressions for the modified potential $V(\phi)$, contingent upon the ``Hidden Higgs Condition" (see Fig.~\ref{fig1}), and the equation of state parameter $w$. Using Eq. (\ref{HiddenHiggs}) in conjunction with Eq. (\ref{HiggsV}) we have:
\begin{equation}V(\phi ) = \epsilon - {\dot \phi}^2/2 
\end{equation}

The Hidden Higgs scenario has a field that climbs {\it up} the potential hill as opposed to classical inflaton field, which rolls down, (See Fig. 2). Note that, the time evolution of the field in Fig. 2 is from right (time zero) to the left (time present). However such behavior for inflaton fields is not without precedent (\cite{Veneziano} and \cite{brax} and references therein). In the current setting, however, a reason for this could be attributed to a force that keeps the Cosmological Higgs fields hidden, which would let the field amass substantial amount of kinetic energy (see Fig. 3), once it reaches the minimum of the potential.  This could happen if, for example, the {\it electroweak Higgs} could fuel the Cosmological Higgs field just before particle production in the early universe. The possibility of new physics coming into play cannot be ruled out if the symmetry is broken after the decay begins. 
\begin{figure}[htb]
\includegraphics[width=0.87\linewidth, height=0.55\linewidth]{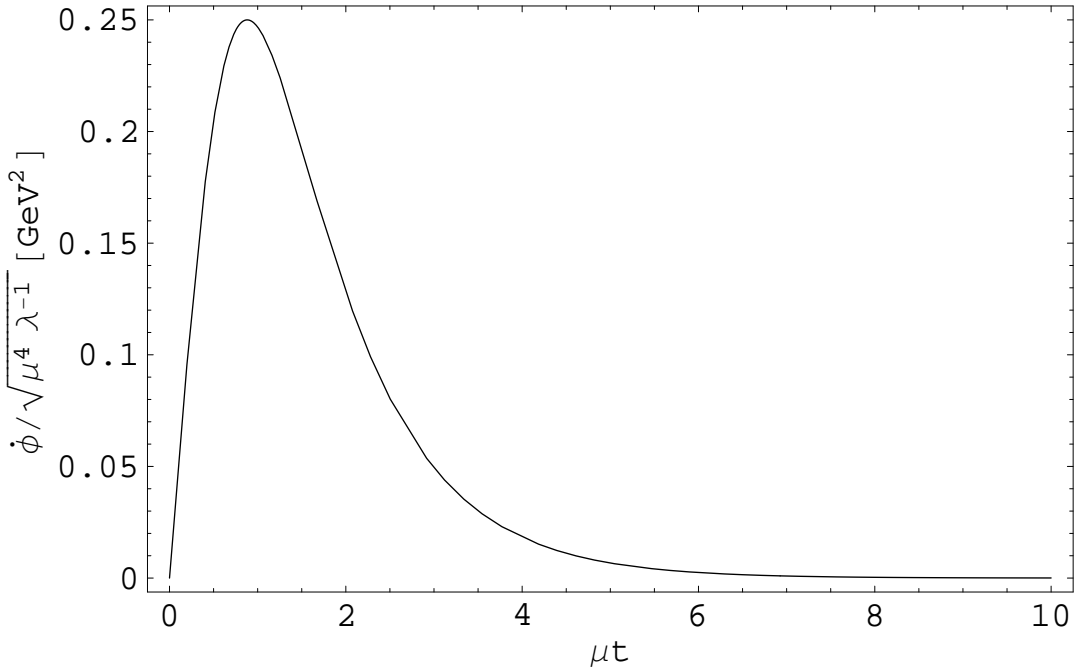}
\caption{The decay rate indicating regimes of change in kinetic energy.}
\label{fig3}
\end{figure}

For a scalar field, it is well known that Equation of State parameter $w = w_{\phi} = p_{\phi}/\rho_{\phi}$ where $p_{\phi}$ and $\rho_{\phi}$ are the pressure and energy density of the dark energy, respectively.
Therefore,
\begin{equation}
 w =  { {\dot \phi}^2/2 - V(\phi) \over
         {\dot \phi}^2/2 + V(\phi) } 
\label{eos}
\end{equation}
Rearranging and simplifying  Eq. (\ref{eos}), using Eq. (\ref{HiddenHiggs}) and Eq. (\ref{HiggsV}) we obtain:
\begin{equation}
w = -1 + {1 \over \epsilon} ( {1 \over 2} {\dot \phi}^2 -
   {\lambda \over 4} \phi^4 + {\mu^2 \over 2} \phi^2 ) 
  = -1 + { {\dot \phi}^2 \over \epsilon } .
\end{equation}
We find that $w$ approaches $-1$ soon after $\phi$ decays appreciably, as ${{\dot \phi}^2 \ll \epsilon}$. This scenario does not predict the value of $\epsilon $, which can be given the observed value of dark energy, ${\approx 10^{-48} GeV^{4}}$, by hand \cite{carroll}. Since $w$ never goes below $-1$, this model does not generate ``phantom energy." Conversely, one can use the lack of observed evolution of $w$ out into the universe to put limits on $\mu$ and $\lambda$.

In conclusion, the energy density of the local universe is dominated by the zero point fluctuations of the Cosmological Higgs field oscillating at a highly stable false vacuum having a nonzero positive expectation value $ \approx \epsilon$, which is quite contrary to classical inflation \cite{linde}, \cite{guth}. We note that the constant zero point fluctuation term in the Higgs potential makes the potential invariant to field translations small compared to $\epsilon$, and there perhaps less sensitive to initial conditions \cite{vanHolten03}.

BP is grateful for useful discussions with Grant Matthews and Chris Kolda during a visit to the University of Notre Dame.

\end{document}